\documentstyle[prd,aps,eqsecnum,twocolumn,floats]{revtex}

\def\beq{\begin{equation}}
\def\eeq{\end{equation}}
\def\beqa{\begin{eqnarray}}
\def\eeqa{\end{eqnarray}}
\def\bfig{\begin{figure}}
\def\efig{\end{figure}}

\input{psfig}

\begin{document}
\fnsymbol{footnote}
\wideabs{

\title{Magnetic effects on the viscous boundary layer damping of the r-modes in neutron stars}

\author{Gregory Mendell}
\address{LIGO Hanford Observatory,
P.O. Box 1970 S9-02, Richland, WA 99352}

\date{\today}
\maketitle
\begin{abstract}
This paper explores the effects that magnetic fields have on the viscous boundary layers (VBLs)
that can form in neutron stars at the crust-core interface, and
it investigates the VBL damping of the gravitational-radiation
driven $r$-mode instability.
Approximate solutions to the magnetohydrodynamic equations valid in the VBL are found for
ordinary-fluid neutron stars.
It is shown that magnetic fields above $10^9$ Gauss significantly
change the structure of the VBL, and
that magnetic fields decrease the VBL damping time.
Furthermore, VBL damping completely suppresses the $r$-mode instability
for $B \gtrsim 10^{12} \, {\rm Gauss}$.
Thus, magnetic fields will profoundly affect the VBL damping of the $r$-mode instability in hot young pulsars
(that are cool enough to have formed a solid crust).
One can speculate that magnetic fields can affect the VBL damping of this
instability in LMXBs and other cold old pulsars
(if they have sufficiently large internal fields).

\end{abstract}
}

\section{Introduction}
\label{sectionI}

Gravitational radiation emitted by the $r$-modes always
tends to drive these modes unstable, while internal dissipation most likely completely suppresses this instablity in all
stars except neutron stars \cite{andersson,fried-morsink,lom,aks,owen-etal}
(see Lindblom \cite{lreview} for a review).  Understanding this instability in neutron stars is currently of
great interest because unstable $r$-modes could
emit gravitational radiation with an intensity that is detectable by gravitational-wave
observatories, such as the proposed enhanced version of LIGO \cite{owen-etal}.
A necessary step in determining the detectability of gravity waves from the $r$-modes
is to find the critical angular velocity that triggers the onset of the instability
as a function of temperature.  This splits the problem into two temperature regimes:
newborn hot young neutron stars \cite{lom,aks,owen-etal},
and cold old neutron stars spun up by accretion, such as in
low-mass x-ray binary systems (LMXBs) \cite{bildsten,akst,levin}.
The reason for this split is that below approximately $10^9 \, {\rm K}$, the interior
of a neutron star is expected to contain regions of superfluid neutrons mixed
with lower concentrations of superconducting protons and normal electrons
\cite{baym-patheck,pines-alpar,sauls,epstein}.
For fluid models, bulk viscosity is the dominant form of dissipation
for $T > 10^9 \, {\rm K}$ \cite{lmo}; while for $T < 10^9 \, {\rm K}$ shear viscosity dominates,
and mutual friction due to electron scattering off quantized neutron vortices probably has little effect
on the $r$-modes \cite{lm2000}.

However, Bildsten and Ushomirsky \cite{bu} showed that when a solid crust is present,
the shear dissipation in the viscous boundary layer (VBL)
that forms at the crust-core interface is by far the most important
suppression mechanism of the $r$-mode instability.
Their work was extended by Andersson {\it et al.} \cite{ajks}, Rieutord \cite{rieutord},
Levin and Ushomirsky \cite{levinushomirsky}, and Lindblom, Owen, and Ushomirsky \cite{lou}.
The state-of-the-art calculation now includes the effects of the Corelois force, angular structure,
equation of state, and compressiblity on the VBL.  However, magnetic effects on the
neutron star VBL have not yet been published.

The purpose of this paper is to explore the effects of magnetic fields on the VBL in neutron stars and
to investigate the VBL damping of the gravitational-radiation
driven $r$-mode instability.
Approximate solutions to the magnetohydrodynamic (MHD) equations valid in the VBL are found for
ordinary-fluid neutron stars.
Neutron stars are expected to form a solid crust for $\rho \lesssim
1.5 \times 10^{14} \, {\rm g/cm}^3$ and for tempertures
below an approximate melting temperature of $T \cong 10^{10} \, {\rm K}$
\cite{douchinhaensel,haensel}.  Once the crust-core region cools
below $10^9 \, {\rm K}$ superfluid effects
could become important.  However, uncertainties exist in the quantum nature of the fluids, and
it is possible either the protons, or the neutrons, or both fluids could be in the normal state at
the location of the VBL (see Epstein \cite{epstein} for a review of some of these uncertainties).
Thus, this paper will focus on the case of an ordinary-fluid core.
The crust is treated as a perfectly rigid,
infinitely conducting solid.

In this paper, it is shown that magnetic fields above $10^9$ Gauss significantly
change the structure of the VBL, and
that magnetic fields decrease the VBL damping time.
Furthermore, VBL damping completely suppresses the $r$-mode instability
for $B \gtrsim 10^{12} \, {\rm Gauss}$.
Magnetic field strengths of hot young pulsars are typically similar to this lower bound.
Cold old pulsars, however, tend to have fields less than $10^{11} \, {\rm Gauss}$.
Thus, assuming interior
fields are of a similar magnitude to external fields,
magnetic fields will profoundly affect the VBL damping of the $r$-mode instability in hot young pulsars
(that are cool enough to have formed a solid crust).
One can speculate that magnetic fields can affect the VBL damping of this
instability in LMXBs and other cold old pulsars
(if they have sufficiently large internal fields).
This possibility, and how superfluidity might effect it, are discussed further
in the conclusion.  Also, the results of
long term magnetic evolution on the interior field
need to be considered when applying the results in this paper \cite{srinietal,ruderman1991}.
Finally, the complications associated with nonlinear magnetic effects, such as the azimuthal
winding of field lines \cite{rls}, should also be taken into account.
Such effects are not likely to change
the general conclusions presented here,
though nonlinear effects are beyond the scope of this paper.

The next section reviews the MHD equations for neutron stars.
Sec.~\ref{sectionII} finds an approximate solution to the MHD equations valid in the VBL.
Sec.~\ref{sectionIII} discusses limits on the magnetic field and the VBL length scales.
Sec.~\ref{sectionIV} presents the results of numerical calculations for the VBL damping times
and the critical angular velocity for two models of the magnetic field.  Checks on the
self-consistency of the results are made at the ends of Secs.~\ref{sectionIII} and \ref{sectionIV}.
Caveats about the approximations used to obtain the results are discussed, and comments on how this work
could be extended, especially to include superfluid effects, are made
in Sec.~\ref{sectionVI}.

\section{MHD Equations}
\label{sectionIb}

The MHD equations for a mixture of superfluid
neutrons, and ordinary-fluid protons and electrons were derived by Easson and Pethick \cite{eassonpethick}.
More recently, the MHD equations for a mixture of superfluid neutrons, type II superconducting protons,
and ordinary-fluid electrons were derived by Mendell \cite{mendell98}, based on the earlier work
of Mendell and Lindblom \cite{mendell-lindblom1991}, Mendell \cite{mendell1991,mendell1991b},
and Lindblom and Mendell \cite{mendell-lind}.  To study the ordinary-fluid case,
the ordinary-fluid limit of the equations in Mendell \cite{mendell98} is taken \cite{eassonpethickfootnote}.
In this limit, scattering between particles causes them to approximately flow as a single fluid.
However, the presence of magnetic fields does allow small but finite electrical currents to exist.  Thus,
the relative velocity between the protons and electrons is not exactly zero, but nearly so.
These currents are accounted for by the MHD approximation.

The ordinary-fluid limit of the equations is accomplished by defining the average
velocity, $\vec{v}$, in terms of the mass densities
and velocities of the neutron, $\vec{v}_n$, and the average velocity of the charged particles
(the protons and electrons), $\vec{u}$, by
\beqa
\rho \vec{v} = \rho_n \vec{v}_n + \rho_p \vec{u},
\eeqa
where $\rho = \rho_n + \rho_p$, and $\rho_n$ and $\rho_p$ are the mass densities of the
neutrons and protons respectively.
For simplicity, terms of order $\rho_e/\rho_p$ are ignored.
The dynamic equation for $\vec{v}$ is found by taking the equations in
Mendell \cite{mendell98} for $\vec{v}_n$ and $\vec{u}$ and performing
this averaging.
Furthermore, ignoring small corrections due to electrical currents, it is a good
approximation to replace the electron velocity with the average velocity in the resulting equations.
Finally, quantized vortices do not occur in ordinary-fluids, and the forces associated
with them are not included.
Thus, for a star rotating uniformly in equilibrium with angular velocity $\Omega$,
the equation for $\vec{v}$ for small perturbations can be written in the corotating frame as
\beqa
\partial_t \delta \vec{v} && + 2\vec{\Omega} \times \delta \vec{v} = - \vec{\nabla} \delta U
\nonumber \\
&& + {1 \over \rho}
\left ( {\delta \vec{J} \over c} \times \vec{B} \right )
+ {1 \over \rho} \vec{\nabla} \cdot (2 \eta \delta \tensor{\sigma}). \label{veqn}
\eeqa
The only other independent vector degree of freedom in the MHD limit is the magnetic field, $\vec{B}$, given
in the corotating frame by
\beqa
\partial_t \delta \vec{B} = \vec{\nabla} \times (\delta \vec{v} \times \vec{B}), \label{Beqn}
\eeqa
\beqa
\vec{\nabla} \cdot \delta \vec{B} = 0. \label{DivBeqn}
\eeqa
Equations~(\ref{veqn})-(\ref{DivBeqn}) are the standard textbook MHD equations,
except the major
contribution to the mass density comes from the neutrons.
Only the dominant dissipative effect due to the shear of the fluid, $\tensor{\sigma}$,
is included.
When magnetic diffusion is important, magnetic effects lead to Hartmann type boundary layers
\cite{ah73}.
However, note that Easson \cite{easson} calculates that the ratio of the magnetic diffusion time scale
to the viscous diffusion time scale in a typical neutron star is roughly $10^{6} (10^{10} \, {\rm K}/T)^4$.
Thus, the magnetic diffusion term has been ignored in Eq.~(\ref{Beqn}).
This means the magnetic effects on the VBL found here are quite different from the Hartmann case.
Finally, note that $\delta U = \delta p /\rho + \delta \Phi$, where $p$ is the pressure
and $\Phi$ is the Newtonian gravitational potential (see \cite{mendell-lind} and references therein).

The MHD limit is valid for studies of oscillations with phase velocities much
less than the speed of light, and frequencies
much less than the plasma and cyclotron frequencies.
Under these circumstances the above equations,
along with the mass conservation laws and equations of state, completely determine
the dynamics of the system.  All the other
vector fields of interest are determined in terms of $\delta \vec{v}$ and $\delta \vec{B}$.
Specifically, for phase velocities much less than the speed of light, the displacement
current can be ignored in Ampere's law, and the current density is given by
\beqa
\delta \vec{J} = {c \over 4 \pi} \vec{\nabla} \times \delta \vec{B}. \label{deltaJeqn}
\eeqa
For an infinitely conducting crust, there will also be a surface current density
at the crust-core interface, given by
\beqa
\delta \vec{I} = {c \over 4 \pi} \Bigl [ \delta \vec{B} \times \hat{r} \Bigr ]_{r  = R_c}, \label{surfI}
\eeqa
where $R_c$ is the radius of the core.  For finite constant conductivity, magnetic fields can diffuse into the
crust via
\beqa
\partial_t \delta \vec{B} = {c^2 \over 4 \pi \Gamma} \nabla^2 \delta \vec{B}, \label{Bdiffuse}
\eeqa
where $\Gamma$ is the conductivity.  However, while Eq.~(\ref{Bdiffuse}) is not too difficult solve,
in this paper it will only be necessary to use this equation to place
bounds on the electron drift velocity and the rate of ohmic dissipation in the crust.
Returning to the discusion of the quantities in the core,
the equilibrium charge density is negligible (as is the perturbed charge density for frequencies
much less than the plasma frequency), and thus the number densities of the electrons and protons
are equal: $n_e = n_p = \rho_p/m_p$.
Thus, the relative velocity between the protons and electrons is given by
\beqa
\delta \vec{v}_p - \delta \vec{v}_e = {m_p \over e \rho_p} \delta \vec{J}. \label{relvpe}
\eeqa
It will be shown near the end of Sec.~\ref{sectionIII} that
the fractional difference between the proton and electron velocity
for the $r$-mode frequency is of the order $10^{-10}$
for typical neutron numbers.  This fact, along with the strong
coupling between the protons and neutrons, justifies the approximation
\beqa
\delta \vec{v}_e \cong \delta \vec{v}_p \cong \delta \vec{v}_n \cong \delta \vec{v}.
\eeqa

Finally, when the above approximations hold and the conductivity is high, electrons
(being the least massive charge carrier)
respond to make the Lorentz force on them negligible.  Thus, again ignoring small corrections
due to the current density, the electric field is given by
\beqa
\delta \vec{E} = - {\delta \vec{v} \over c} \times \vec{B} - {\vec{v} \over c} \times \delta \vec{B}.
\eeqa

\section{Approximate Viscous Boundary Layer Solutions}
\label{sectionII}

Approximate solutions to the MHD equations are
found in this section.
The equilibrium magnetic field is assumed to be arbitrary (for now) except that it is static in the corotating
frame, and it is restricted such that no equilibrium electrical currents exist.
This implies the equilibrium structure of the star is unchanged
by the presence of the magnetic field.
To facilitate the manipulation of tensor quantites, a rotating
spherical coordinate basis will be used, and indices will be raised and lowered using the metric
tensor.  In this basis, the equilibrium velocity is $v^a = \Omega \phi^a$.
Note that, following the notation of previous studies,
Latin indices are space indices, except where it is understood that
$n$, $p$, $e$, and $c$
refer to neutrons, protons, electrons,
and the crust respectively.

Let $\delta v^a$ and $\delta U$ describe
the nondissipative, lowest order (when expanded in powers of $\Omega$), $r$-mode solution
ignoring magnetic fields. These solutions are valid in
the bulk of the core where viscous
and magnetic forces are small compared to the Coreolis force.
However, because of viscosity,
the no-slip boundary condition must be applied to the velocity at the crust-core interface.
This causes a VBL to form.
In the VBL, the magnitudes of the viscous, magnetic, and Coreolis forces become comparible. Thus, all these
forces have effects on the structure of the VBL.
Let $\delta \tilde{v}^a$ and $\delta \tilde{U}$ describe
the corrections that must be added to the standard r-mode solutions to enforce
the no-slip boundary condition
\beqa
\Bigl[ \delta v^a + \delta \tilde{v}^a \Bigr ]_{r = R_c} = 0. \label{noslip}
\eeqa
Finally, note that for the approximation of an infinitely conducting crust,
the only boundary
condition that must be applied to the magnetic field is that it
gives the surface current density given by Eq.~(\ref{surfI}).
The surface current density is small, and does not significantly change the no-slip
boundary condition given by Eq.~(\ref{noslip}) (see near the end of Sec.~\ref{sectionIII}).
In this case, no magnetic diffusion takes place, and the details of what happens in the
crust can be ignored.

It is possible to find approximate equations for the corrective quantities
by taking the following steps.
First, let all perturbed quantities in the corotating frame have time dependence ${\rm exp}(i\kappa \Omega t)$.
Second, note that
since $\delta v^r$ vanishes everywhere, the boundary condition that the radial
components of the velocity fields vanish at the crust-core boundary is satisfied if the
radial component, $\delta \tilde{v}^r$, also vanishes.
Finally, following the work of Lindblom, Owen, and Ushomirsky \cite{lou}, and references therein,
all angular derivatives of the corrective quantities are ignored.
This is valid, since the no-slip boundary condition forces the
corrective velocity to rapidly change in the radial direction.
Thus, terms involving radial derivatives dominate all others.
(For certain angles, certain caveats must be added to this statement.
These angles are discussed in Secs.~\ref{sectionIII} and \ref{sectionVI}.)
Typically, the ratio of radial derivatives to angular derivatives is the
ratio of the radius of the core to the boundary-layer thickness.
Taking these steps, the resulting equations (expressed in a corotating spherical coordinate basis) are
\beqa
\delta \tilde{B}^a = - {i \over \kappa \Omega} B^r \partial_r \delta \tilde{v}^a, \label{deltaBeqn}
\eeqa
\beqa
&& \partial_r \delta \tilde{U} =
i \, {V_{\rm A}^2 \over \kappa \Omega}{B_r \over B}
\left ( {B^\theta \over B} \partial_r^2 \delta \tilde{v}_\theta
+ {B^\phi \over B} \partial_r^2 \delta \tilde{v}_\phi \right ) \nonumber \\
&& \qquad + 2 \Omega r {\rm sin}^2\theta \delta \tilde{v}^\phi
, \label{partialrUeqn}
\eeqa
\beqa
&& i\kappa\Omega \delta \tilde{v}^\theta - 2 \Omega {\rm cos}\theta ({\rm sin}\theta \delta \tilde{v}^\phi) =
{\cal F} \partial_r^2 \delta \tilde{v}^\theta , \label{Vthetaeqn}
\eeqa
\beqa
&& i\kappa\Omega ({\rm sin}\theta \delta \tilde{v}^\phi) + 2 \Omega {\rm cos}\theta \delta \tilde{v}^\theta =
{\cal F} \partial_r^2 ({\rm sin}\theta \delta \tilde{v}^\phi) , \label{Vphieqn}
\eeqa
where
\beqa
{\cal F} = -i\left [{V_{\rm A}^2 \over \kappa \Omega} \left ( {B_r \over B} \right )^2
+ {i \eta  \over \rho} \right ] . \label{calFeqn}
\eeqa
In Eqs.~(\ref{partialrUeqn}) and (\ref{calFeqn}) $V_{\rm A}^2$ is the square of the Alfv\'{e}n wave speed, defined as
\beqa
V_{\rm A}^2 \equiv {B^2 \over 4 \pi \rho}, \label{VAeqn}
\eeqa
and throughout this paper note that
$B = |\vec{B}|$.
Note that the magnetic effects on the VBL vanish if the radial component of the equilibrium
field, $B_r$, vanishes.
Also, Eq.~(\ref{partialrUeqn}) for the quantities $\delta \tilde{U}$
is included for completeness, but plays no further role in this paper.
Finally, Mendell \cite{mendell98} has shown that Alfv\'{e}n waves are replaced by
cyclotron-vortex waves in a superfluid neutron, type II superconducting proton,
ordinary-fluid electron plasma.
Thus, it is possible to speculate on what happens when quantum fluids
are present.  This will be done in Sec.~\ref{sectionVI}, where expected
complications with the superconducting-superfluid case will also be discussed.

To find solutions to the equations, first note that the $\phi$-component of the equations becomes
identical to the $\theta$-component when
\beqa
\pm i {\rm sin}\theta \delta \tilde{v}^\phi = \delta \tilde{v}^\theta .  \label{vthetaphieqn}
\eeqa
The corrective boundary layer solution is then found by allowing all perturbative quantities
to vary as ${\rm exp}[ik(R_c - r)]$.
It is then easy to show that solutions exist for
\beqa
k_\pm = K_\pm \sqrt{ {\Omega
\over
{V_{\rm A}^2 \over \kappa \Omega}\left ( {B_r \over B} \right )^2 + {i \eta \over \rho}}} , \label{kopmeqn}
\eeqa
where
\beqa
&& K_\pm = \sqrt{\kappa \pm 2{\rm cos} \theta} . \label{Koeqn}
\eeqa
Choosing solutions where ${\rm Im}(k_\pm) \ge 0$, so that the corrective quantities
decay exponentially as $r \rightarrow 0$, the general solution
for the $\theta$-component of the velocity is
\beqa
\delta \tilde{v}^\theta = [ C_+ e^{ik_+ (R_c - r)}
+ C_- e^{ik_- (R_c - r)} ] e^{i \kappa \Omega t} .
\eeqa
The $\phi$-component of velocities follows from Eq.~(\ref{vthetaphieqn}).

If we apply the no-slip boundary condition given in Eq.~(\ref{noslip})
the constants $C_\pm$ are given by
\beqa
C_\pm = -{1 \over 2}(\delta v^\theta \pm i {\rm sin}\theta \delta v^\phi). \label{Cpmeqn}
\eeqa
The components of the standard $r$-mode velocity
in the corotating frame (with their time dependence cancelled) are \cite{lou}
\beqa
\delta v^\theta = -i A r^{m - 1} {\rm sin}^{m-1}\theta e^{im \phi}, \label{Vrmodetheta}
\eeqa
\beqa
{\rm sin}\theta \delta v^\phi = A r^{m - 1} {\rm sin}^{m-1}\theta {\rm cos} \theta e^{im \phi} . \label{Vrmodephi}
\eeqa

As shown in previous papers, the viscous damping rate is given by integrating the shear over the core:
\beqa
{1 \over \tau_v} = {1 \over 2 E}\int 2 \eta \delta \sigma^*_{ab} \delta \sigma^{ab} d^3x, \label{oneovertauv}
\eeqa
where $E$ is the energy of the mode as defined in e.g., Lindblom, Mendell, and Owen \cite{lmo},
but limited to the core.
The largest contribution to the integral comes from the radial derivatives of
the corrective boundary layer velocity, so that
\beqa
\delta \sigma^*_{ab} \delta \sigma^{ab} = {1 \over 2} R_c^2 (|\partial_r \delta \tilde{v}^\theta|^2
+ |\partial_r ({\rm sin}\theta \delta \tilde{v}^\phi)|^2), \label{shearsquared}
\eeqa
ignoring terms smaller than these by a factor of the boundary layer thickness over the core radius.
Thus, it can be shown that the solutions presented in this section give a VBL damping time of
\beqa
\tau_v = {2 \pi \over \eta {\cal I}} {2^{m + 3} (m + 1)! \over m (2m + 1)!!}
\int_0^{R_c} \rho \left ( {r \over R_c} \right )^{2m + 2} dr,
\label{tauv}
\eeqa
where
\beqa
&& {\cal I} = \int_0^{2\pi} \int_0^\pi \Bigr [ |k_+|^2 d_+ (1 - {\rm cos}\theta)^2 \nonumber \\
&& \qquad \qquad \qquad + |k_-|^2 d_- (1 + {\rm cos}\theta)^2 \Bigl ] {\rm sin}^{2m -1} \theta d\theta d\phi.
\label{Im}
\eeqa
Using the definition
\beqa
k_\pm \equiv {2 \pi \over \lambda_\pm} + { i \over d_\pm}, \label{lambdaandddefined}
\eeqa
the boundary layer thicknesses in Eq.~(\ref{Im}) and
boundary layer oscillation wavelengths are defined by
\beqa
\lambda_\pm \equiv {2 \pi \over {\rm Re}(k_\pm)}, \label{lamdefined}
\eeqa
\beqa
d_\pm \equiv {1 \over {\rm Im}(k_\pm)}. \label{ddefined}
\eeqa
The results given here then reduce to those given by Lindblom, Owen, and Ushomirsky \cite{lou}
in the limit $B \rightarrow 0$.

\section{Bounds on the magnetic field and boundary layer length scales}
\label{sectionIII}

In this section, numerical estimates are made for the characteristic length
scales that determine the properties of the VBL.  A lower bound is placed
on $B$, above which magnetic field effects are dominant.  Towards these ends,
let $B_r = B$ for this section.

The magnetic effects on the VBL are determined by examining Eq.~(\ref{kopmeqn}).
In this equation, if the magnetic term is larger than the viscous term, then
magnetic effects on the VBL length scales will be important. Thus, the condition for
this to happen is
\beqa
{V_{\rm A}^2 \over \kappa \Omega} \ge {\eta \over \rho}. \label{VAlimiteqn}
\eeqa
Defining $\rho_{1.5e14} = \rho/(1.5 \times 10^{14} \, {\rm g} \cdot {\rm cm}^{-3})$,
$T_{10}  = T/(10^{10} \, {\rm K})$ (and so on for other numeric subscripts in cgs units thoughout
the rest of this paper) the ordinary-fluid viscosity is given by \cite{cutler-lind}
\beqa
\eta = \biggl (2.73 \times 10^{14} {{\rm g} \over {\rm cm} \cdot {\rm s}} \biggr ) \rho_{1.5e14}^{9/4} T_{10}^{-2}.
\label{etao}
\eeqa
Using Eqs.~(\ref{VAeqn}) in Eq.~(\ref{VAlimiteqn})
yields the following lower bound on the magnetic field, such that magnetic effects dominate
the VBL properties:
\beqa
B \ge (4.6 \times 10^9 \,{\rm G}) \kappa^{1/2} \Omega_{2000\pi}^{1/2} \rho_{1.5e14}^{9/8} T_{10}^{-1}.
\label{Bordinarylimit}
\eeqa
Note that hot young ordinary-fluid neutron stars have external fields much larger
than the lower bound given above.
Thus, magnetic fields will profoundly affect the VBLs in these stars if the internal fields
are of a similar magnitude.
Furthermore, note that the range of values of $B$ in LMXBs probably lie on either side of the
lower bound given above.  Thus, it is uncertain to what extent magnetic
effects on the VBL are important in these stars or other cold old pulsars with
weak magnetic fields.  Further complications with interpreting the results presented here
in cold old pulsars occur due
to the likelyhood that the protons and neutrons condense into superconducting
and superfluid states.  It has already been pointed out that
Alfv\'{e}n waves are replaced by
cyclotron-vortex waves in that case \cite{mendell98}.
More is said about how this might change the results in this paper for
cold old pulsars in Sec.~\ref{sectionVI}.

The characteristic length scales of the VBL are determined by examining Eq.~(\ref{kopmeqn})
and the definitions of the length scales given in Eqs.~(\ref{lamdefined}) and (\ref{ddefined}).
Dropping the ``$\pm$'' symbol for the purposes of this discussion, since
the VBL damping rate is controlled by terms of
the form $\eta |k|^2 d = \eta (4\pi/\lambda^2 + 1/d^2)d$, it would seem that the smallest length scales
would determine the VBL damping rate. However, the ratio $d/\lambda$ turns out to be more important
when magnetic effects are important.  This point will
be elaborated on in the next section.
Also, only real frequencies, $\kappa \Omega$, will be considered.  This is valid either when the system is driven
at a real frequency, or when the imaginary part of the frequency is small and can be ignored,
except near certain special angles.
The latter situation is the case for the r-modes and most other neutron star oscillations
of interest.
In this case, the coefficient $K_\pm$ is either purely real or purely imaginary depending on the
angle $\theta$.  The more complicated case that occurs near the special angles already mentioned
is discussed at the end of this section and in Sec.~\ref{sectionVI}.
Also, to further simplify the discussion, it will always be assumed that $\kappa > 0$.

Taylor expanding Eq.~(\ref{kopmeqn}) for the case when $B$ is less than
the lower bound given in Eq.~(\ref{Bordinarylimit}), the viscous term dominates, and the
boundary layer thicknesses and wavelengths are given to lowest order by
\beqa
d_{\eta} = {\lambda_{\eta} \over 2\pi} = {1 \over |K_\pm|} \sqrt{{2 \eta \over \Omega \rho}} . \label{detaoeqn}
\eeqa
For typical neutron star numbers these length scales are
\beqa
d_{\eta} = {\lambda_{\eta} \over 2\pi} = {0.024 \,{\rm cm} \over |K_\pm|} \rho_{1.5e14}^{5/8} T_{10}^{-1}
\Omega_{2000\pi}^{-1/2}. \label{valdetaoeqn}
\eeqa
These equations agree in form with the standard result,
but differ from previous results by numerical factors of
order unity that are incorporated into $|K_\pm|$.

Taylor expanding Eq.~(\ref{kopmeqn}) for the case when $B$ is greater than
the lower bound given in Eq.~(\ref{Bordinarylimit}),
the magnetic terms dominate, and the wave number is given by
\beqa
k_\pm = K_\pm \sqrt{{\kappa \Omega^2 \over V_{\rm A}^2}}
\biggl [1 - {i \over 2} {\eta \over \rho} {\kappa \Omega \over V_{\rm A}^2} \biggr ].
\label{Taylorkpmo}
\eeqa
The problem now divides in two cases
corresponding to when $K_\pm$ is real and imaginary respectively.

For the case of real $K_\pm$, the boundary layer length scales are given to lowest order by
\beqa
{\lambda_{B} \over 2 \pi} = {1 \over |K_\pm| \kappa^{1/2}}
{V_{\rm A} \over \Omega}, \label{lamBoeqn}
\eeqa
\beqa
d_{B/\eta} = {2 \over |K_\pm| \kappa^{3/2}}
{V_{\rm A}^3 \rho \over \Omega^2 \eta}.
\eeqa
The subscripts indicate whether these quanties depend on purely magnetic, or a ratio of magnetic to viscous
quantities.  Substituting in values for the parameters gives
\beqa
{\lambda_{B} \over 2 \pi} = {0.017 \,{\rm cm} \over |K_\pm| \kappa^{1/2}}B_{4.6e9}\Omega_{2000\pi}^{-1}\rho_{1.5e14}^{-1/2},
\label{vallamBoeqn}
\eeqa
\beqa
d_{B/\eta} = {0.033 \,{\rm cm} \over |K_\pm| \kappa^{3/2}}B_{4.6e9}^{3}\Omega_{2000\pi}^{-2}\rho_{1.5e14}^{-11/4} T_{10}^{2}.
\label{valdBetaoeqn}
\eeqa
Obviously $\lambda_{B}$ and $d_{B/\eta}$ do not appear much different
from what is given in Eq.~(\ref{valdetaoeqn}) for the limiting field
of $B = 4.6 \times 10^9 \,{\rm Gauss}$.  However, for a typical $10^{12} \,{\rm Gauss}$ field
\beqa
{\lambda_{B} \over 2 \pi} = {3.7 \,{\rm cm} \over |K_\pm|
\kappa^{1/2}}B_{12}\Omega_{2000\pi}^{-1}\rho_{1.5e14}^{1/2},
\label{vallamB12oeqn}
\eeqa
\beqa
d_{B/\eta} = {3.4 \times 10^5 \,{\rm cm} \over |K_\pm|
\kappa^{3/2}}B_{12}^{3}\Omega_{2000\pi}^{-2}\rho_{1.5e14}^{-11/4} T_{10}^{2}.
\label{valdB12etaoeqn}
\eeqa
Note that in Eq.~(\ref{valdB12etaoeqn}) the boundary layer thickness is approaching $30\%$ the size of the star.
Thus, the approximation
that $d/R_c$ is small becomes less valid as the size of $B$ increases.
For the case of imaginary $K_\pm$, the roles of $\lambda$ and $d$ become interchanged. Thus, for this case,
the results are the same as for Eqs.~(\ref{lamBoeqn}) - (\ref{valdB12etaoeqn}) with
\beqa
&& d \rightarrow {\lambda \over 2 \pi} , \nonumber \\
&& \lambda \rightarrow 2 \pi d.
\eeqa

Two checks on the self-consistency of the solution can now be made.
First, it can be shown that for the length scales presented here that
the fractional difference between the proton and electron velocities is neglibible in the VBL,
as has been assumed.  Substituting Eq.~(\ref{deltaBeqn}) into Eq.~(\ref{deltaJeqn}) and the result into
Eq.~(\ref{relvpe}) gives
\beqa
&& \delta \tilde{v}_p^a - \delta \tilde{v}_e^a =  i \left ( {e B_r \over m_p c} \right )
\left ( {m_p^2  c^2 k^2 \over 4 \pi \kappa \Omega \rho_p e^2} \right ) \nonumber \\
&& \qquad \times \biggl ({ \delta \tilde{v}^\theta \over {\rm sin} \theta }  \phi^a
- {\rm sin} \theta \delta \tilde{v}^\phi \theta^a \biggr ) ,
\eeqa
which for typical neutron star numbers gives
\beqa
&& \delta \tilde{v}_p^a - \delta \tilde{v}_e^a  = i {[(1.3 \times 10^{-10} \, {\rm cm}^2) k^2]
B_{12} \over \kappa \rho_{p13} \Omega_{2000\pi}  }
\nonumber \\
&& \qquad  \times \left ( {B_r \over B} \right )
\biggl ( { \delta \tilde{v}^\theta \over {\rm sin} \theta }  \phi^a
- {\rm sin} \theta \delta \tilde{v}^\phi \theta^a \biggr ) .
\eeqa
Thus, the fractional difference between these velocities is small as long
as $(10^{-10} \, {\rm cm}^2) k^2 \ll 1$, which is true for the length scales presented in this section.
Second, at the boundary the electrons can slip to produce the surface current
density given by Eq.~(\ref{surfI}). The surface current will actually be spread
out over a thickness equal to the skin-depth, $\zeta$, that the magnetic
field penetrates the crust.  Examining Eq.~(\ref{Bdiffuse}), it can be seen that for the $r$-modes
that $\zeta$ is given by
\beqa
\zeta = \sqrt{c^2 \over 2 \pi \kappa \Omega \Gamma}. \label{skindepth}
\eeqa
The conductivity, $\Gamma$, is given by \cite{eassonpethick},
\beqa
\Gamma = 2 \times 10^{25} \, {\rm s}^{-1} \rho_{p13}^{3/2}T_{10}^{-2}, \label{conductivity}
\eeqa
and thus,
\beqa
\zeta = {3.4 \times 10^{-5} \, {\rm cm} \over \sqrt{\kappa}}
\Omega_{2000\pi}^{-1/2} \rho_{p13}^{-3/4} T_{10}. \label{valskindepth}
\eeqa
Within the skin-depth region of the crust $\delta \tilde{v}_e^a = - (m_p/e\rho_p) \delta \tilde{J}^a$ and
$\delta \tilde{J}^a \cong \delta \tilde{I}^a/\zeta$ \cite{bittenfootnote}.  Thus, the electron contribution to
the average velocity at $r = R_c$, which is $(\rho_e/\rho) \delta \tilde{v}_e^a$,
is negligible as long as
$(10^{-10} \, {\rm cm}^2)(\rho_e/\rho) (k/\zeta) \ll 1$, which also holds
true for the length scales presented here.
Ohmic dissipation within the skin-depth region is considered at the end of the next
section, once the dominant VBL length scales are determined.

Finally, consider the special case that occurs near certain special angles, defined by
\beqa
\kappa \pm 2 {\rm cos}X_\pm = 0. \label{thetaA}
\eeqa
For regions of the crust-core boundary where $\theta$ is near $X_\pm$, the
small imaginary part of $\kappa$ becomes important.
To understand this, make the replacement $\kappa \rightarrow \kappa + i/(\Omega \tau_v)$, and note that
$\Omega \tau_v \gg 1$ for the situations of interest.
In this case, for $\theta = X_\pm$ the
angular derivative $\partial k_\pm / \partial \theta$ becomes proportional to
$\sqrt{\Omega \tau_v/i}$, which is large (and infinite at $X_\pm$ when the imaginary part
of $\kappa$ is ignored).
However, these regions
make little contribution to the VBL damping rate,
since $|k|^2 d \propto 1/\sqrt{\Omega \tau_v}$ is small near $X_\pm$.
Thus corrections to the solutions near $\theta = X_\pm$ are not made in this paper, nor
have they been made in previous studies.

\section{Boundary Layer Damping Times and the Critical Angular Velocity}
\label{sectionIV}

As shown in previous studies, gravitational radiation emitted by the r-modes always
tends to drive these modes unstable.  Of primary interest to this study is how magnetic fields change the onset of
the instability of the r-modes via their effect on the VBL damping rate. Since the VBL
damping rate dominates all other forms of dissipation, the onset occurs when this rate equals the
gravitational-radiation growth rate.  Thus, we can calculate the critical angular velocity for the onset
of the gravitational-radiation instability by solving
\beqa
\tau_v = \tau_{GR}, \label{tauvEQtaugr}
\eeqa
where $\tau_v$ is given by Eq.~(\ref{tauv}).
It has been shown that $\tau_{GR}$ for the $r$-modes is given by
\beqa
\tau_{GR} = \tilde{\tau}_{GR} \left ( {\Omega_o \over \Omega} \right )^{2m + 2} , \label{taugr}
\eeqa
where $\Omega_o  = \sqrt{\pi G \bar{\rho}}$, and $\bar{\rho}$ is the average density of the star.

Attention is now restricted to the case used in previous studies: the $m = 2$ r-mode for a $1.4 M_\odot$ $n = 1$
polytrope, as described in, e.g., Lindblom, Owen, and Mendell \cite{lom}, and Lindblom and Mendell \cite{lm2000}.
Adoption of this model allows easy comparison with previous studies. (Note also that the $m = 2$ case is the
most susceptible to the $r$-mode instability.)
For this case, $\kappa_0  = 2/3$, the stellar radius is $12.53 \, {\rm km}$, and $\Omega_o = 8413 \,{\rm s}^{-1}$.
Note that the
maximum angular velocity, where mass shedding occurs at the equator, is roughly $2\Omega_o/3$ (for any equation of state).
The density at the crust-core boundary is given approximately by
$1.5 \times 10^{14} \,{\rm g/cm}^3$ (see \cite{douchinhaensel,haensel}).
Using this density, the characteristic gravitational-radiation growth time
is $\tilde{\tau}_{GR} = 4.25 \,{\rm s}$ \cite{lou}, and
the core radius is $R_c = 11.01 \, {\rm km}$.
The viscosity is given in Eq.~(\ref{etao}).

Two models of the magnetic field will be considered.
For both models the overall magnitude of the equilibrium magnetic field, $B$, is taken
to be a constant in the boundary layer.  The models are defined by
\beqa
&& {\rm Model \,\,\,\, I} \qquad \qquad B_r = B = {\rm constant} , \nonumber \\
&& {\rm Model \,\, II} \qquad \qquad B_r = B {\rm cos}\Theta ,
\eeqa
where $\Theta$ is the angle between the magnetic axis and the vector to the point $(\theta, \phi)$.
This is given in terms of the location of this point by
\beqa
{\rm cos}\Theta = {\rm sin} \Lambda {\rm sin}\theta{\rm cos}\phi + {\rm cos}\Lambda {\rm cos}\theta, \label{cosTheta}
\eeqa
where the magnetic axis is chosen to lie in the $xz$ plane of the corotating
frame, and $\Lambda$ is the angle between this axis and the rotation axis.
Note that the $\theta$ and $\phi$
components of the magnetic field cannot be zero for Model II if $B$ is constant.
However, these components do not enter into the effects on the VBL.

Now note that the VBL damping time, $\tau_v$, is a function of the temperature,
angular velocity, and magnetic field, but in general the dependence on these quantities
cannot be factored out of
Eq.~(\ref{tauv}) as they have been in previous studies (but see towards the end of this section).
Still, one can numerically evaluate
$\tau_v$ for various temperatures and models of
the magnetic field.  The results are presented in Table~\ref{tableI}
for $\Omega = \Omega_o$.  Note that the values of $\tau_v$ agree
with the results of Lindblom, Owen, and Ushomirsky \cite{lou} in the $B = 0$ case.

\begin{table}
\caption{The VLB damping time in seconds for $\Omega = \Omega_o$.
Model II numbers are for $\Lambda = \pi/2$. \label{tableI}}
\begin{tabular}{ccccc}
B &$\tau_v$&$\tau_v$&$\tau_v$&$\tau_v$\\
$\,$&Model I &Model I &Model II &Model II \\
$\,$&$T = 10^{8} \,{\rm K}$ &$T = 10^{10} \,{\rm K}$&$T = 10^{8} \,{\rm K}$ &$T = 10^{10} \,{\rm K}$\\
\tableline
0.0 &51.89 &5189 &51.89 &5189 \\
$10^9$ &51.89 &5056 &51.89 &5141 \\
$10^{10}$ &51.88 &1603 &51.89 &2706 \\
$10^{11}$ &50.57 &161.0 &51.41 &303.4 \\
$10^{12}$ &16.03 &16.10 &27.06 &30.37 \\
\end{tabular}
\end{table}

The temperature dependence of the critical angular velocity is presented in
Figs.~\ref{FigOrdB1flds}-\ref{FigOrdB2angs}.
In these figures, the horizontal dashed line corresponds to $2\Omega_o/3$,
which is the approximate maximum angular velocity for which mass-shedding occurs.
Thus, the regions above this
line are unphysical and are kept only to illustrate the dependence of the curves for the range
of magnetic field magnitudes typically found in neutron stars.
Also, the curves probably only apply for $T > 10^9 \,{\rm K}$, since only
ordinary-fluid neutron stars are considered in this paper. They have been
extended below this value to make comparison with other studies easier.  This also makes
allowances for uncertainties in the superfluid transition temperature at the crust-core boundary.

\bfig \centerline{\psfig{file=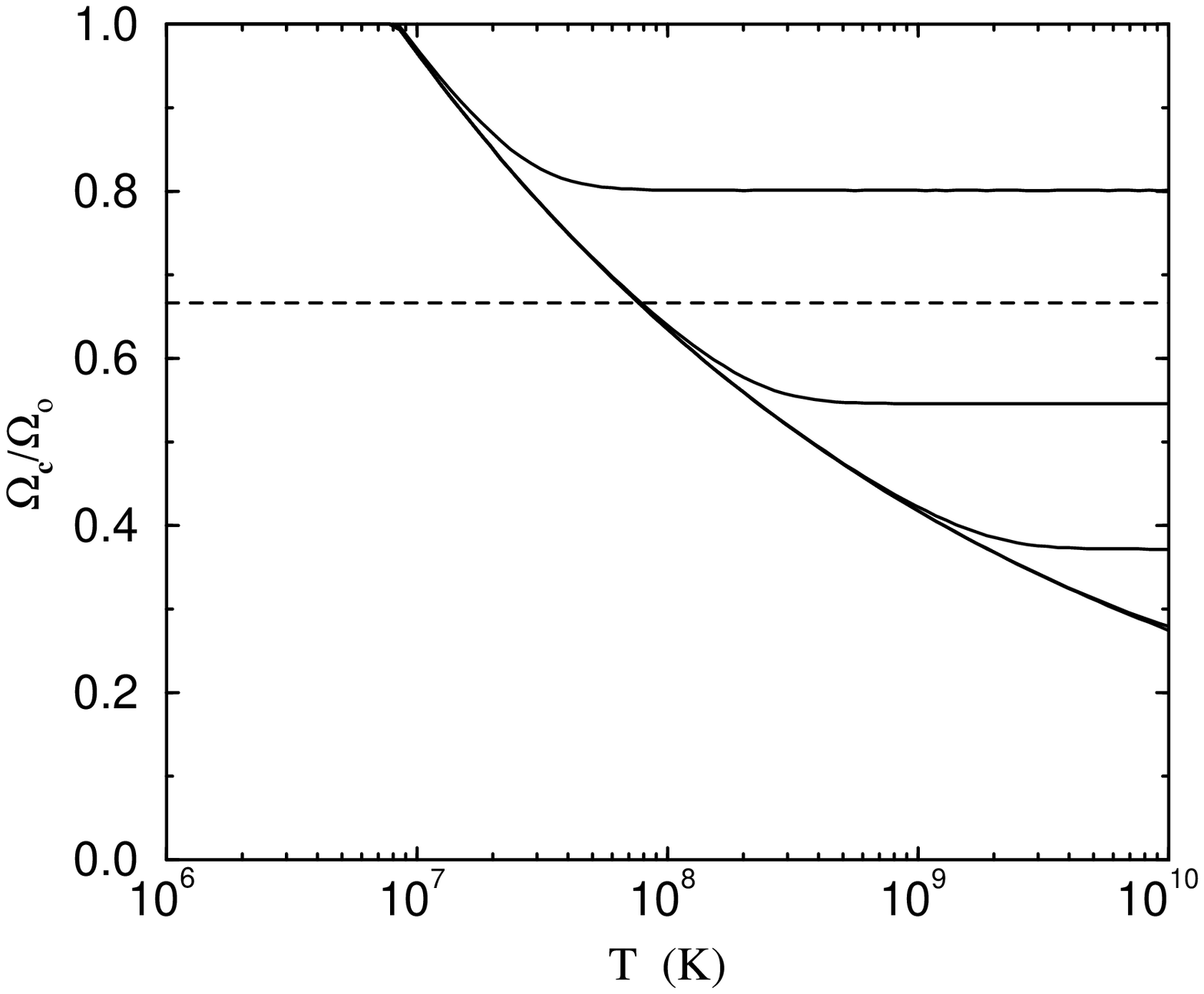,height=2.4in}} \vskip 0.3cm
\caption{Temperature dependence of the critical angular velocities
for Model I and $B = 0$, $10^{9}$ (indistinguishable from $B = 0$), $ 10^{10}$, $10^{11}$,
and $10^{12}$ Gauss, from bottom to top.\label{FigOrdB1flds}} \efig

\bfig \centerline{\psfig{file=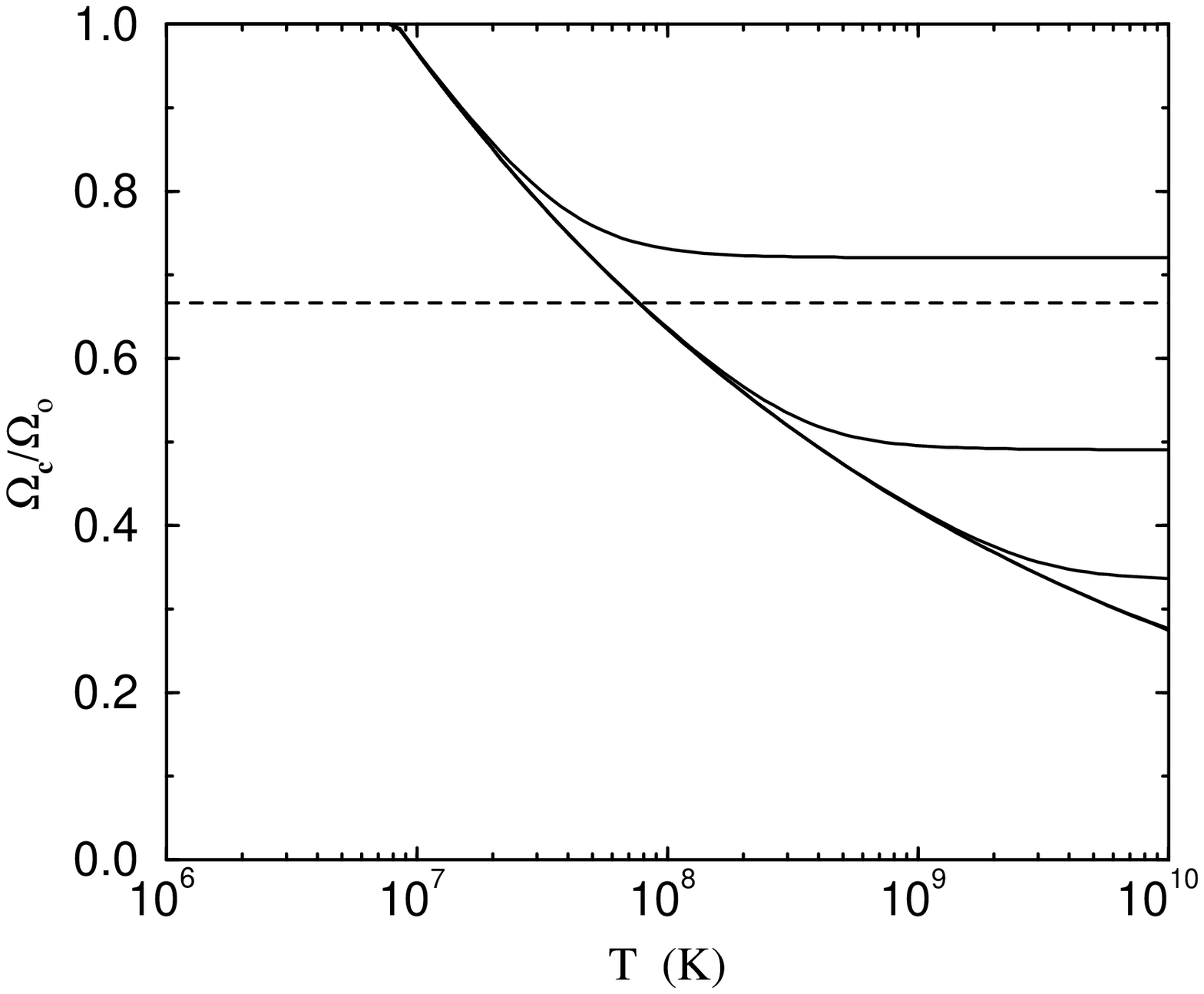,height=2.4in}} \vskip 0.3cm
\caption{Temperature dependence of the critical angular velocities
for Model II and $B = 0$, $10^{9}$ (indistinguishable from $B = 0$), $ 10^{10}$, $10^{11}$,
and $10^{12}$ Gauss, from bottom to top. \label{FigOrdB2flds}} \efig

\bfig \centerline{\psfig{file=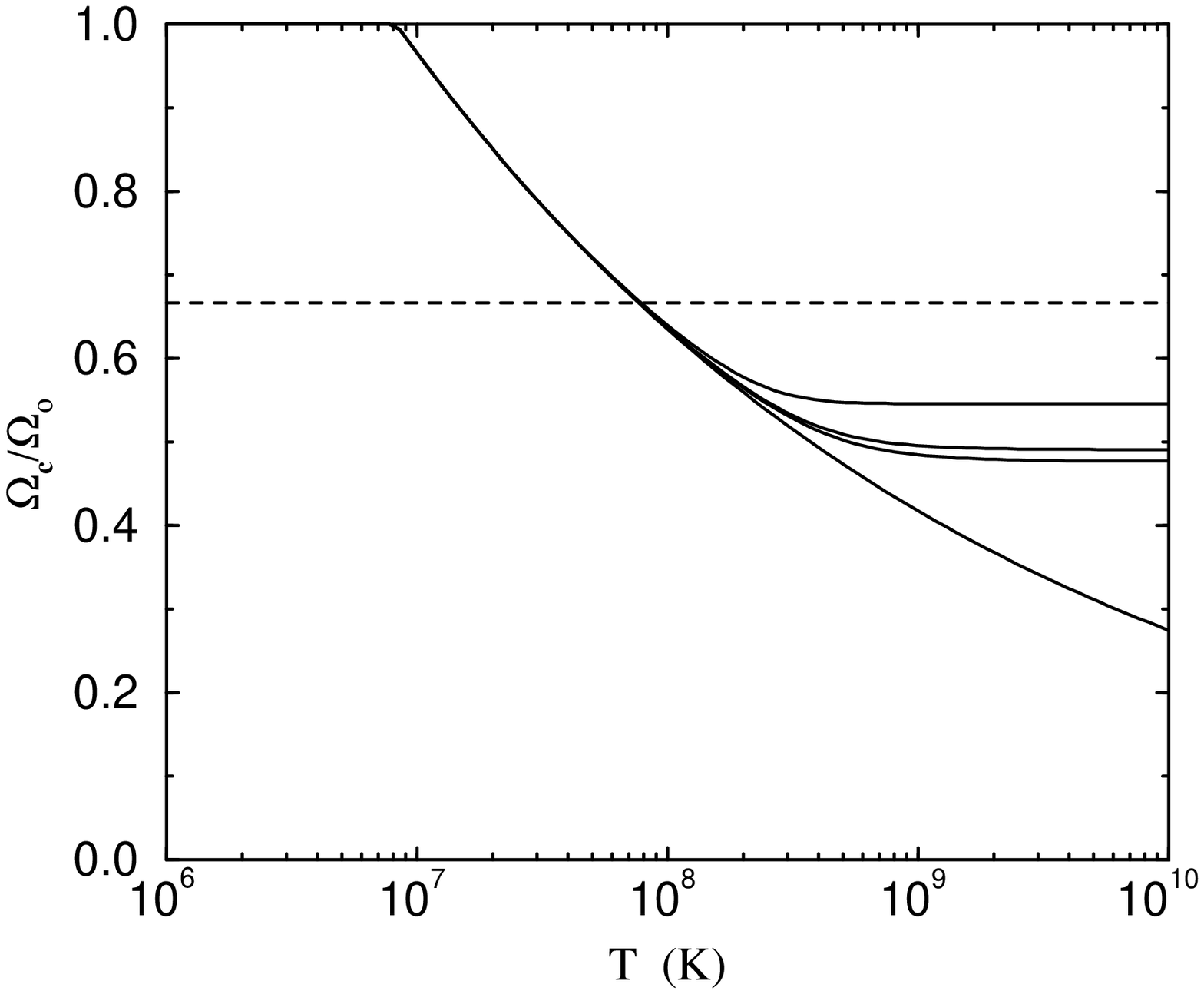,height=2.4in}} \vskip 0.3cm
\caption{Temperature dependence of the critical angular velocities
for Model I $B = 0 \,{\rm Gauss}$ (bottom curve), Model I $B = 10^{11} \,{\rm Gauss} $ (top curve), and
Model II $B = 10^{11} \,{\rm Gauss}$ for $\Lambda = 0$ and $\Lambda = \pi/2$ (middle two curves).
\label{FigOrdB2angs}} \efig

\bfig \centerline{\psfig{file=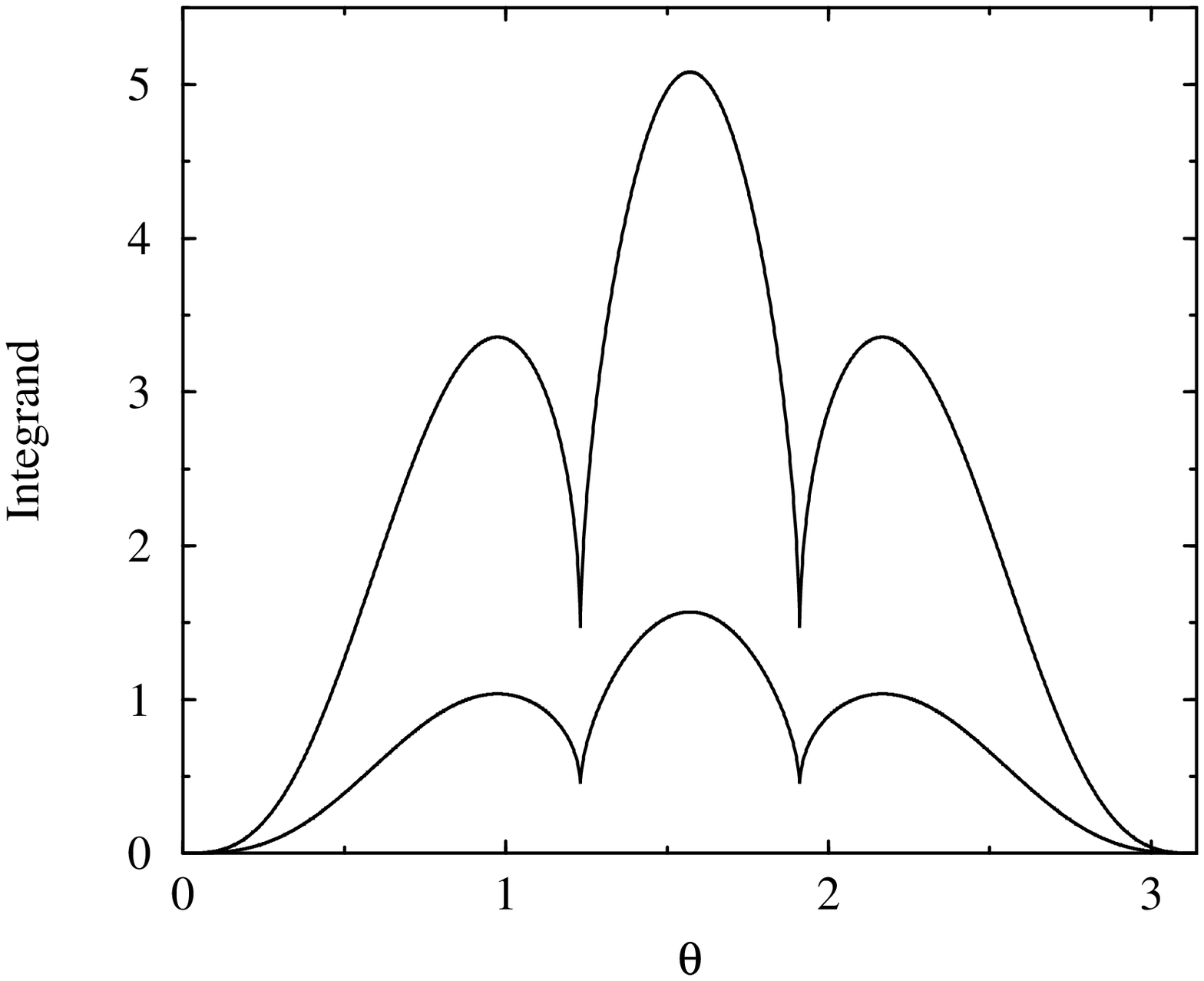,height=2.4in}} \vskip 0.3cm
\caption{Integrand that gives ${\cal I}$ for $\phi = 0$, $T = 10^8 \, {\rm K}$, $\Omega = \Omega_o$,
and $B = 0$ (bottom curve) and Model I $B = 10^{12} \, {\rm Gauss}$ (top curve).
\label{FigSupIntsB00andB12}} \efig

Figure~\ref{FigOrdB1flds} shows the
temperature dependence of the critical angular velocities
for Model I.
Figure~2 shows a similar result for Model II when
$\Lambda = \pi/2$.  Figure~3 compares the $B = 0$ case with the Model
I and  Model II ($\Lambda = 0$ and $\pi/2$) $B = 10^{11} \,{\rm Gauss}$ cases.
(Results for other values of $\Lambda$ fall between the two Model II curves shown.)  As
would be expected, the angular dependence of the magnetic field for Model II reduces the overall effect
of the magnetic field on the critical angular velocity, as compared with Model I.
Overall, it is seen that for magnetic fields $B \gtrsim 10^{12} \, {\rm Gauss}$
that the $r$-mode instability is completely suppressed in ordinary-fluid neutron stars.

Note that the critical angular velocity
becomes temperature independent for high temperatures.
This can be understood by considering the length scales from the last section.
It is apparent that for large $B$ that the VBL damping rate varies as
\beqa
{ 1 \over \tau_v} \sim \eta |k|^2 d \sim \eta {d_{B/\eta} \over \lambda_{B}^2},
\eeqa
which implies for large $B$ that $\tau_v$ scales as
\beqa
\tau_v = 16.10 \, {\rm s} \,B_{12}^{-1}, \label{valtauvB12}
\eeqa
independent of temperature or angular velocity [see Eqs.~(\ref{etao}), (\ref{vallamBoeqn})-(\ref{valdBetaoeqn})].
The constant of proportionality was found by fitting the data in the table
for Model I.
Substituting this equation into Eq.~(\ref{tauvEQtaugr}) and using Eq.~(\ref{taugr}) gives
the following results for Model I critical angular velocities
\beqa
{\Omega_c \over \Omega_o} = 0.80 B_{12}^{1/6}. \label{valOmegacritB12}
\eeqa
The Model II result is identical to this,
but with a coeffient of 0.72.  This equation gives the
exact numerical results for the regions where the critical angular
velocity curves flatten out.

Thus, the temperature independence of the critical angular velocity for high
fields is completely explained by the length-scale analysis done here.
However, before accepting this result, one needs to check
that the small imaginary part of the frequency does not significantly alter
it, since the imaginary part of $k_\pm$ is also becoming small when the above equation
holds true.  Consider the replacement $\kappa \rightarrow \kappa + i/(\Omega\tau_v)$
in Eq.~(\ref{Taylorkpmo}): the imaginary part of $\kappa$ can
be ignored as long as
\beqa
\tau_v >  {1 \over \kappa^2} {\rho \over \eta} {V_{\rm A}^2 \over \Omega^2}.
\eeqa
Thus, the results presented here are self-consistent for
\beqa
\tau_v >  (9.3 \, {\rm s}) B_{12}^2 T_{10}^2 \left ( {\Omega_o \over \Omega} \right )^2.
\label{Boundtauo}
\eeqa

Finally, now that $d_{B/\eta}$ is seen as the relevant thickness of the
VBL when magnetic fields are important, one can
check that the ohmic dissipation rate within the skin-depth region of the crust
is smaller than the shear dissipation rate within the VBL.  Note that the ohmic
disspipation rate scales as
$(\zeta/\Gamma) |\delta \tilde{J}|^2$.  For finite
conductivity, the magnetic field at the boundary is
continuous, and thus the derivative of the magnetic field in the crust, which gives
the current,
is proportional to $(c/4\pi)\delta \tilde{B}/\zeta$.  Thus, the ohmic dissipation rate scales as
$[1 / (\zeta\Gamma)] |c \delta \tilde{B}/(4 \pi)|^2$, and
$\delta \tilde{B} \sim [Bk/(\kappa\Omega)]\delta \tilde{v}$.
On the other hand, the shear dissipation rate in the VBL
scales as $\eta |k \delta \tilde{v}|^2 d_{B/\eta}$.
Substituting in typical numbers, the ratio of the ohmic to shear dissipation rates
is $10^{-6} B_{12}^{-1} T_{10} \sqrt{\Omega/\Omega_o}$
[see Eqs.~(\ref{etao}), (\ref{valdB12etaoeqn}),(\ref{skindepth})-(\ref{valskindepth})],
which justifies the neglect of ohmic dissipation in this paper.

\section{Conclusions}
\label{sectionVI}

While perhaps not obvious at first, it should now be clear why the results found in this paper hold true.
It has been shown that magnetic fields increase the thickness of the VBL, which would tend to increase
the VBL damping time. However, vibrations
of the magnetic field lines, corresponding to Alfv\'{e}n waves,
produce a larger shear than one would expect due to the VBL thickness alone.
The wavelength of these vibrations is the distance the Alfv\'{e}n waves travel in one period
of the r-mode oscillation.  This is a longer distance than
viscous diffusion can travel in the same amount of time.
Thus, the shear is smaller than when the magnetic field is zero, but because the VBL thickness is increased,
the shear acts over a larger volume.  The net result is that magnetic fields decrease the VBL damping time.

Before concluding this paper,
some caveats need to be made about the validity of the approximations made in this paper.
First, the assumption that angular derivatives are small needs some further comment.
The last figure show the integrand that gives ${\cal I}$ for $\phi = 0$ for two cases.
This integrand determines the
VLB damping rate. For the model used in the last section, the special
angles discussed in Sec.~\ref{sectionIII}, in radians, are located at $X_- = 1.23$ and $X_+ = 1.91$.
It can be seen that the approximation that angular derivatives are small breaks down near $X_\pm$.
However, correcting this
should not greatly affect the results in this paper, since
these regions make small contributions the VBL damping rate.
However, future studies could consider correcting for the angular behavior near $X_\pm$.
Second, future studies could consider
cases where the imaginary part of the frequency influences
the structure of the VBL, as when the bound in Eq.~(\ref{Boundtauo}) is violated.
Third, as noted after Eq.~(\ref{valdB12etaoeqn}), the approximation $d/R_c \ll 1$ breaks down
for $B \gtrsim 10^{12} \, {\rm Gauss}$. Future studies could consider extending the results presented
here to the case of fields larger than this bound.
More importantly, future studies should
explore the superconducting-superfluid case.
Mendell \cite{mendell98} has shown that Alfv\'{e}n waves are replaced by
cyclotron-vortex waves in a superfluid neutron, type II superconducting proton, ordinary-fluid electron
plasma.  Since the square of cyclotron-vortex wave velocity scales
linearly with $B$, one would speculate that the results
for superfluid neutron stars would be similar to those presented in
Eqs.~(\ref{valtauvB12}) and (\ref{valOmegacritB12})
but with $\tau_v \propto B^{-1/2}$
and $\Omega_c/\Omega_o \propto B^{1/12}$, for large $B$.
However, there are complications with applying the no-slip boundary condition when both
the neutrons and protons form inviscid quantum condensates.  In this case,
in principle, both the neutrons and protons can slip at the crust-core boundary.
The electrons can slip as well, since the crust is conducting.  Instead,
the boundary conditions are determined by pinning effects.  In the strong pinning limit
neither the neutrons or the protons can slip at the boundary.  The pinning
of proton vortices between neutron vortices, the possible expulsion of proton
vortices during the spindown phase of a pulsar, and mutual friction effects
due to the scattering of electrons off quantized neutron and proton vortices, are further
effects that need to be considered. Thus, future work on the correct boundary conditions
is needed before the superfluid case can be worked out.

In conclusion, this paper has shown that magnetic fields profoundly affect the VBL in neutron stars.
Note that this paper treated the crust as perfectly rigid, while crustal motion increases the VBL damping rate
by a factor between $1$ and $400$ \cite{levinushomirsky}.  Thus, one should treat the results presented here as
upper bounds on the critical angular velocity.  (However, other effects not yet considered,
like turbulance, could again increase the critical angular velocity.)
Still, it is clear that magnetic fields of the magnitude typically found in neutron
stars can substantially shorten the VBL damping time and, if sufficiently large, can suppress the $r$-mode
instability.

\acknowledgments I wish to thank L.~Lindblom for helpful discussions concerning
this work.  This research was supported by NSF grant PHY-9900767.



\begin{references}
\bibitem{eassonpethickfootnote} Note that Easson and Pethick
do not include equations for the superfluid neutrons, making this limit harder to take in their case.  They
also include relativistic corrections for the motion of the electrons, and the anisotropy of the
viscosity in a magnetic field.  For simplicity, these corrections are ignored in this paper.
\end{references}
\end{document}